\documentclass[aps,pra,twocolumn,groupedaddress,showpacs]{revtex4}

\usepackage{graphicx}
\usepackage{dcolumn}
\usepackage{bm}
\usepackage{verbatim}
\usepackage{amsmath,amssymb}
\usepackage{color}

\begin{document}
\title{System of classical nonlinear oscillators as a
coarse-grained quantum system}
\author{Milan Radonji\' c}
\author{Slobodan Prvanovi\' c}
\author{Nikola Buri\' c}
\email[]{buric@ipb.ac.rs}
\affiliation{Institute of Physics, University of Belgrade,
Pregrevica 118, 11080 Belgrade, Serbia}

\begin{abstract}
Constrained Hamiltonian dynamics of a quantum system of nonlinear
oscillators is used to provide the mathematical formulation of a
coarse-grained description of the quantum system. It is seen that
the evolution of the coarse-grained system preserves constant and
minimal quantum fluctuations of the fundamental observables. This
leads to the emergence of the corresponding classical system on a
sufficiently large scale.
\end{abstract}


\maketitle

\section{Introduction}

Relation between quantum and classical mechanics (QC-relation) is
a very complex one with many complementary facets. Over the years,
since the discovery of the quantum theory, many different and more
or less related aspects of the QC-relation have been investigated.
The studied problems could be artificially divided into two main groups.
The first group entails the problems of formal or mathematical relations
between quantum and classical formalisms (an excellent review is
\cite{Landsman}). Problems of the other group are related to the
description of physical reasons or processes that effectuate the quantum
to classical transition \cite{decoh1,decoh2,Ian}. Our goal in this paper
is to explore yet another formal QC-relation and interpret it as the
mathematical formulation of a coarse-graining that is necessary in the
quantum to classical transition.

Comparison of typical formal features of classical and quantum
mechanics is facilitated if the same mathematical framework is
used in both theories. It is well known, since the work of Kibble
\cite{Kibble1,Kibble2,Kibble3}, that the quantum evolution,
determined by the linear Schr\"{o}dinger equation, can be
represented using the typical language of classical mechanics,
that is as a Hamiltonian dynamical system on an appropriate phase
space, given by the Hilbert space geometry of the quantum system.
This line of research was later developed into the full geometric
Hamiltonian representation of quantum mechanics \cite{Heslot,Weinberg1,
Weinberg2,Weinberg3,Hughston1,Hughston2,Hughston3,Hughston4}.
Such geometric formulation of quantum mechanics has inspired natural
definitions of measures of the entanglement \cite{Hughston4},
and has been used to model the spontaneous collapse of the state vector
\cite{Hughston,Adler}.

It has been realized recently that the geometric formulation of quantum
mechanics provides particularly suitable framework for discussions of nonlinear
constraints that might be imposed on a quantum system
\cite{AnnPhys,Hughston_sym,Hughston_met}. In particular, it was shown in
reference \cite{AnnPhys} that a quantum system of two qubits constrained to
be always in the manifold of separable states shows the characteristic
qualitative features of classical Hamiltonian dynamical systems, that can not
be realized by the unconstrained Schr\"odinger evolution. The idea is further
explored in \cite{PhysA} and applied to a general spin system, i.e.\ to a
quantum system with a finite Hilbert space. Study of the
QC-relations for such systems is hampered by the fact that there
is no classical mechanical model which after quantization gives
the quantum system.

In this paper we consider systems based on the Heisenberg $H_4$
dynamical algebra, say a collection of oscillators possibly
nonlinear and interacting. Such a system is quantized to give the
quantum system of oscillators. Our main result is that the quantum
system of oscillators constrained with a specific type of
constraints is equivalent to a finite dimensional Hamiltonian
system that preserves constant and minimal quantum fluctuations of
the fundamental observables during the entire evolution. This
Hamiltonian system is close to the classical one if some classicality
parameter is small. Finally we shall propose an interpretation of these
formal results as the mathematical formulation of the emergence of
classical systems from a coarse-grained description of quantum
systems.

The paper is organized as follows. Geometric Hamiltonian
formulation of quantum mechanics and in particular the quantum
constrained evolution for a general quantum system is formulated
in section 2. In section 3, that contains our main result, this
formalism is applied to study the evolution of a system of
quantum oscillators with particular constraints. In section 4 a
complete construction of the classical system based on the
constrained quantum system is presented. Section 5 contains a
discussion and an interpretation of the formal results from sections 3 and 4.

\section{Hamiltonian formulation of constrained quantum dynamics}

\subsection{Hamiltonian framework for quantum systems}

Consider a quantum system with separable and complete Hilbert space ${\cal
H}$. Schr\"odinger dynamical equation on ${\cal H}$ generates a
Hamiltonian dynamical system on an appropriate symplectic manifold. The
symplectic structure, which is needed for the Hamiltonian formulation of the
Schr\"odinger dynamics, is provided by the imaginary part of the
unitary scalar product on ${\cal H}$. In fact the Hilbert space
${\cal H}$ is viewed as a real manifold ${\cal M}$ with a complex
structure, given by a linear operator $J$ such that $J^2=-1$. If
${\cal H}$ is finite n-dimensional then ${\cal M}\equiv
\mathbf{R}^{2n}$, but in general ${\cal M}$ is an infinite dimensional
Euclidean manifold. Real coordinates $\{(x_i,y_i),\> i=1,2,\dots\}$ of a
point $\psi\in {\cal H}\equiv {\cal M}$ are introduced using
expansion coefficients $\{c_i,\> i=1,2,\dots\}$ in some basis
$\{|i\rangle,\> i=1,2,\dots\}$ of ${\cal H}$ as follows
\begin{subequations}
\label{e:psi}
\begin{align}
&|\psi\rangle=\sum_i c_i|i\rangle,\quad c_i=\frac{x_i+i y_i}{\sqrt{2}},\\
&x_i=\sqrt{2}\,\Re(c_i),\quad
y_i=\sqrt{2}\,\Im(c_i),\quad i=1,2,\dots
\end{align}
\end{subequations}
Alternatively, if ${\cal H}$ is identified with some space of functions
$L_2(\mathbf{R}^N)$ with $q\in \mathbf{R}^N$ then the real and
imaginary parts of $\psi(q)=(\phi(q)+i\>\!\pi(q))/\sqrt{2}$ give two real fields $(\phi(q),\pi (q))$ representing
the coordinates of the real infinite manifold ${\cal M}$.

Besides the complex structure $J$, the real manifold ${\cal M}$
has Riemannian and symplectic structure. Since ${\cal M}$ is real,
it is natural to decompose the unitary scalar product on ${\cal H}$ into it's
real and imaginary parts
\begin{equation}
\langle\psi_1|\psi_2\rangle={1\over 2\hbar} G(\psi_1,\psi_2) +
{i\over 2\hbar} \Omega(\psi_1,\psi_2).
\end{equation}
It follows that $G$ is Riemannian metric on ${\cal M}$ and that
$\Omega$ is symplectic form on ${\cal M}$. Furthermore, $J$, $G$ and
$\Omega$ satisfy $G(\psi_1,\psi_2)=\Omega(\psi_1,J\psi_2)$ so that
the space ${\cal M}$ is in fact a K\"ahler manifold. Thus the
manifold ${\cal M}$ associated with the Hilbert space ${\cal H}$
can be viewed as a phase space of a Hamiltonian dynamical system.
A vector $|\psi\rangle$ from ${\cal H}$, associated with a pure quantum
state $\psi $, is represented by the corresponding point $X_{\psi}$ in the
phase space ${\cal M}$. It is convenient to add an abstract
index $a=1,2,\dots$ to the points from ${\cal M}$ like $X^a_{\psi}$ and
to assume the standard summation convention over repeated abstract indices.
On the other hand, summation over coordinate indices $i$, $j$ like in
(\ref{e:psi}) or integration over the argument $q$ in $\phi(q)$,
$\pi(q)$ will always be written explicitly. In all following formulas we shall
set $\hbar=1$.

In the coordinates $(x_i,y_i)$ the Riemannian and the symplectic structures
of ${\cal M}$ are given by
\begin{equation}
G=\begin{pmatrix}\mathbf{1}&\mathbf{0}\cr
\mathbf{0}&\mathbf{1}\end{pmatrix},
\end{equation}
\begin{equation}
\Omega=\begin{pmatrix} \mathbf{0}&\mathbf{1}\cr
-\mathbf{1}&\mathbf{0} \end{pmatrix},
\end{equation}
where $\mathbf{0}$ and $\mathbf{1}$ are zero and unit matrices of
dimension equal to the dimension of the Hilbert space. In the coordinates
$(\phi(q),\pi(q))$ the analogous formulas are
\begin{equation}
G(\psi_1,\psi_2)=\int dq (\phi_1(q)\phi_2(q)+\pi_1(q)\pi_2(q))
\end{equation}
\begin{equation}\label{e:Omega_q}
\Omega(\psi_1,\psi_2)=\int dq(\phi_1(q)\pi_2(q)-\pi_1(q)\phi_2(q)).
\end{equation}
Thus, coordinates $\{(x_i,y_i)$, $i=1,2\dots\} $ or
$\{(\psi(q),\pi(q))$, $q\in {\mathbf R}^N\}$ represent canonical
coordinates of a Hamiltonian dynamical system. Consequently, the
Poisson bracket between two functions $F_1$ and $F_2$ on ${\cal
M}$ corresponding to the symplectic form $\Omega$ is in the
canonical coordinate representation $(\phi(q),\pi(q))$ are given by
\begin{equation}
\{F_1,F_2\}=\int dq ({\delta F_1\over\delta \phi(q)}{\delta
F_2\over\delta\pi(q)}-{\delta F_2\over\delta \phi(q)}{\delta
F_1\over\delta\pi(q)}).
\end{equation}

A one parameter family of unitary transformations on ${\cal H}$
generated by a self-adjoined operator ${\hat A}$ is represented on
${\cal M}$ by a flow generated by the Hamiltonian vector field
$\Omega(-J\hat{A}\psi,\cdot)=(dA)(\cdot)$ with the Hamilton's function
\begin{equation}\label{e:Apsi}
A(X_{\psi})=\langle\psi|\hat A|\psi\rangle\equiv\langle\hat A\rangle_\psi.
\end{equation}
Thus, quantum observables $\hat A$ are represented by functions of the form
$\langle\hat A\rangle_\psi$. Such and only such Hamiltonian flows with the
Hamilton's function of the form (\ref{e:Apsi}) also generate isometries of
the Riemannian metric $G$. More general Hamiltonian flows on ${\cal M}$,
corresponding to the Hamilton's function which are not of the form
(\ref{e:Apsi}), do not generate isometries and do not have the physical
interpretation of quantum observables. In what follows we shall often use the
short-hand notation $A(X_\psi)\equiv A$ and $\langle\hat A\rangle_\psi\equiv \langle\hat A\rangle$, implicitly assuming relation to the state $\psi$.
Not every such function has an interpretation as a classical variable.
If it does then we shall denote it by the corresponding small letter
$a\equiv A\equiv\langle\hat A\rangle$.

It can be seen easily that the Poisson bracket of two Hamilton's functions
relates to the commutator between corresponding observables
\begin{equation}
\{A_1,A_2\}=\langle[\hat A_1,\hat A_2]\rangle.
\end{equation}
The Schr\"odinger evolution generated by Hamiltonian $\hat H$
\begin{equation}
 |\dot\psi\rangle=-i \hat H|\psi\rangle,
\end{equation}
is equivalent to the Hamilton's equations on ${\cal M}$
\begin{equation}
 \dot X_{\psi}^a=\Omega^{ab}\nabla_b H(X_{\psi}).
\end{equation}
In the canonical coordinates $(x_i,y_i)$ the Schr\"odinger evolution is given
by
\begin{equation}
\dot x_i={\partial H\over\partial y_i},\quad \dot y_i=-{\partial
H\over\partial x_i},\quad i=1,2,\dots,
\end{equation}
or in $(\phi(q),\pi(q))$ coordinates by
\begin{equation}
\dot \phi(q)={\delta H\over \delta \pi(q)},\quad \dot
\pi(q)=-{\delta H\over \delta \psi(q)},\quad q\in {\mathbf R}^N.
\end{equation}

We have constructed the Hamiltonian dynamical system corresponding
to the Schr\"odinger evolution equation on ${\cal H}$. In fact,
phase invariance and arbitrary normalization of the quantum states
imply that the proper space of pure quantum states is not the
Hilbert space used to formulate the Schr\"odinger equation, but the
projective Hilbert space. This also is a K\"ahler manifold and can
be used as a phase space of a completely geometrical Hamiltonian
formulation of quantum mechanics. Nevertheless, we shall continue
to use the formulation in which points of the quantum phase space
are identified with the vectors from ${\cal H}$ since it is sufficient for our
main purpose.

\subsection{Constrained quantum systems}

The Hamiltonian framework for quantum dynamics enables one to
describe the evolution of a dynamical system generated by the
Schr\"odinger equation with quite general additional constraints
\cite{AnnPhys,Hughston_sym,Hughston_met}. Suppose that the evolution
given by the Hamiltonian $H$ is further constrained
onto a submanifold $\Gamma$ of ${\cal M}$ given by
 a set of $k$ independent functional equations
\begin{equation}\label{e:Constr}
f_l(X)=0, \> l=1,2,\dots,k.
\end{equation}

Equations of motion of the constrained system are obtained using
the method of Lagrange multipliers. In the Hamiltonian form, the method
assumes that the dynamics on $\Gamma$ is determined by the following set
of differential equations
\begin{equation}\label{e:Xdot}
\dot X=\Omega(\nabla X,\nabla H_{tot}), \qquad H_{tot}=H+\sum_{l=1}^k
\lambda_lf_l,
\end{equation}
that should be solved together with the equations of the constraints
(\ref{e:Constr}). Other approaches to realize the constraints are
possible \cite{Hughston_met}, but the resulting system is not explicitly of
Hamiltonian form. Notice that the total Hamilton's function $H_{tot}$ need
not be given as the quantum expectation of a linear operator on ${\cal H}$.
The Lagrange multipliers $\lambda_l$ are functions on ${\cal M}$ that are
to be determined from the following, so called compatibility, conditions
\begin{equation}\label{e:fdot}
\begin{split}
0=\dot f_l&=\Omega(\nabla f_l,\nabla H_{tot})\\&= \Omega(\nabla
f_l,\nabla H)+\sum_{m=1}^k\lambda_m \Omega(\nabla f_l,\nabla f_m)
\end{split}
\end{equation}
on the constrained manifold $\Gamma$.

There is standard Dirac's approach to the constrained classical Hamiltonian
dynamics \cite{Dirac,Klader}. We shall not go into the it's details that stress
on the distinction between the first and the second class constraints. In order
to apply the standard procedure, the constraints have to be regular. A set of
constraints is irregular if there is et least one such that the derivative of the
constraint with respect to at least one of the coordinates is zero in at least
one point on the constrained manifold. Otherwise the constraints are regular.
In our case the constraints are regular if for all $l$
\begin{equation}\label{e:Reg}
{\delta f_l\over \delta\phi(q)}\neq 0,\quad {\delta f_l\over
\delta\pi(q)}\neq 0,
\end{equation}
for all $q\in {\mathbf R}^N$ and everywhere on the constrained
manifold. If this is not satisfied the Dirac's classification is blurred and the
straightforward application od Dirac's recipe is not possible. It will turn out
that the case of interest here involves precisely the irregular constraints that
cannot be easily replaced by an equivalent set of regular constraints.

We shall now briefly recapitulate the main steps of the general analysis of the
constrained dynamics. The equation (\ref{e:fdot}) can be satisfied in two
fundamentally different ways. First, if the matrix of Poisson brackets
$\{f_l,f_m \}=\Omega(\nabla f_l,\nabla f_m)\equiv
(\mathbf{\Omega}_f)_{l,m}$ computed on $\Gamma$ is nonsingular,
then the multipliers are uniquely determined from
\begin{equation}
\lambda_l=\sum_{m=1}^k (\mathbf{\Omega}_f^{-1})_{l,m}
\Omega(\nabla f_m,\nabla H).
\end{equation}
 The equations of motion (\ref{e:Xdot}) assume the form
\begin{equation}
\dot X=\Omega(\nabla X,\nabla H)+\sum_{l,m=1}^k f_l\>\!
(\mathbf{\Omega}_f^{-1})_{l,m}\Omega(\nabla f_m,\nabla H)
\end{equation}
and should be solved together with the constraints (\ref{e:Constr}). In this
case all the constraints (\ref{e:Constr}) are called primary and of the
second class. In this case $\Gamma$ is symplectic manifold with the
symplectic structure determined by the so called Dirac-Poisson brackets
\begin{equation}
\{F_1,F_2\}_D=\{F_1,F_2\}+\sum_{l,m=1}^k
\{f_l,F_1\}(\mathbf{\Omega}_f^{-1})_{l,m}\{f_m,F_2\}.
\end{equation}

Very different situation occurs if all of the Poisson brackets $\{f_l,f_m \}$
and $\{f_m,H\}$ are zero on the constrained manifold $\Gamma$ and the
regularity condition (\ref{e:Reg}) is trivially satisfied. In this case the
constraints are said to be of the first class. The compatibility conditions do
not specify the multipliers and the constrained dynamics is not uniquely
determined. Nevertheless, once a system with regular first class constraints and
the Hamiltonian $H_{tot}$ is put onto the constrained manifold, the
system remains on that manifold whatever choice is made for the Lagrange
multipliers. Different choices of the multipliers must be considered as leading to
the same physical situation.

Let us stress that the described scheme can be applied and leads to the above
conclusions only if the constraints are regular. If this is not the case then it
might be necessary to fix some or all of the multipliers even if the constraints
appear to be of the first class. The system analysed in the next section is
precisely of this type.

If some of the compatibility equations do not contain multipliers, than for
that condition $\dot f_l=\{f_l,H\}=0$ represents an additional
constraint. These are called secondary constraints, and they must
be added to the system of original constraints (\ref{e:Constr}).
They could be of the first or of the second class. If this enlarged set
of constraints is functionally independent one can repeat the
procedure. At the end one either obtains a contradiction, in which case
the original problem has no solution, or one obtains appropriate
multipliers $\lambda_l$ that need not be uniquely determined.

\section{Dynamics of a quantum system of oscillators with constraints}

The Hilbert space ${\cal H}=L_2(R^{n})$ is the unique irreducible
representation space of the canonical commutation relations given
by the $n-$terms direct sum of Heisenberg $H_4$ algebras. Up to the
normalization and the global phase invariance, this Hilbert space
is the state space of a collection of $n$ quantum oscillators. The
fundamental observables of such a system are represented by $2n$
operators $(\hat Q_i,\hat P_i),\> i=1,2,\dots n$, satisfying
$[\hat Q_i,\hat P_j]=i\delta_{i,j}$ on a dense domain in ${\cal H}$.
The symplectic phase space ${\cal M}$ of the Hamiltonian formulation
of the quantum oscillators system is given as the product of $n$
infinite dimensional symplectic spaces. The canonical coordinates
of this infinite dimensional symplectic space can be written using
the continuous index as: $\phi(q_1,\dots,q_n),\pi(q_1,\dots,
q_n)$ ($q_i\in {\mathbf R}$) or using discrete indices as
$(x^l_i,y^l_i)$ ($l=1,2,\dots n,\> i=1,2,\dots$). A Hermitian
operator $\hat A$ is in the Hamiltonian formulation represented
as function $A(X_{\psi})=\langle\psi|\hat A|\psi\rangle$ on
${\cal M}$. In particular, fundamental observables $\hat Q_i$,
$\hat P_j$ give $2n$ fundamental variables as functions on the
infinite quantum phase space ${\cal M}$, which we shall denote
as $q_i=\langle\hat Q_i\rangle$, $p_i=\langle\hat P_i\rangle$.
The Poisson brackets of the infinite phase space ${\cal M}$ between
the fundamental variables $q_i$, $p_j$ are given by the general
formula (\ref{e:Omega_q}) as
\begin{equation}
\{q_i,p_j\}_{\cal M}=\delta_{i,j},\quad i,j=1,2,\dots n,
\end{equation}
where we stress by the subscript ${\cal M}$ that the Poisson bracket
is computed on the infinite manifold ${\cal M}$, for example as in
(\ref{e:Omega_q}). Notice that the quantum variables of the oscillator
system are represented as functions of the fundamental variables of the
infinite phase space ${\cal M}$ ($x^l_i$, $y^l_i$ or the canonical
fields $\phi(q_1,\dots,q_n)$, $\pi(q_1,\dots,q_n)$) but most of them
can not be represented as functions only of the fundamental
variables $q_i$, $p_j$. A nonlinear operator expression in terms of
$\hat Q_i$, $\hat P_i$ is represented as a function of $x^l_i$, $y^l_i$
($l=1,2,\dots,n$, $i=1,2,\dots$) or functional of $\phi(q_1,\dots q_n)$,
$\pi(q_1,\dots q_n)$, but can not be written as function only of $q_i$,
$p_j$ ($i,j=1,2,\dots,n$). Such expressions involve terms that represent
quantum fluctuations of the fundamental observables, i.e.\ contain the
second or higher order moments, for example fluctuations $(\Delta \hat Q_i)^2
=\langle\hat Q_i^2\rangle-q_i^2$, $(\Delta \hat P_i)^2=\langle\hat P_i
\rangle^2-p_i^2$ and correlations $\langle\hat P_i\hat Q_i+\hat Q_i\hat P_i
\rangle-2 p_i q_i$. Of course, these are functions on ${\cal M}$ but can not be
presented as functions only of $q_i$, $p_j$ ($i,j,=1,2,\dots,n$). A polynomial
expression of $\hat Q_i$, $\hat P_j$ thus involves a function of $q_i$, $p_j$
{\it plus additional terms involving the correlations}. The important observation is
that the correlations can become arbitrary large during typical Schr\"odinger evolution.
However, there is an important exception. Namely, {\it when the system is in the
coherent state}, all moments of $\hat Q_i$ ($\hat P_i$) of order higher than
two are expressible solely in terms of $q_i$ and $\Delta \hat Q_i$ ($p_i$ and
$\Delta \hat P_i$), while the correlations $\langle \hat Q_i^m\hat P_i^n+
\hat P_i^n\hat Q_i^m\rangle-2 q_i^m p_i^n$ ($m,n\in \mathbf N$) vanish.

Previous discussion suggests that a closed dynamical system expressed solely
in terms of the fundamental variables $q_i$, $p_j$ could be obtained from the
quantum system if the Schr\"odinger evolution is additionally constrained to
appropriate coherent state manifold, i.e.\ to preserve constant and minimal
values of the fluctuations of the fundamental observables $\hat Q_i$,
$\hat P_j$. The formalism of constrained quantum Hamiltonian system sketched
in the previous section is ideally suited for the analysis of such systems.
However, as we shall see, the construction of the most appropriate set of
constraints and the analysis thereof is not straightforward. In this section we
deal with the construction of the constrained system. Physical interpretation of
the constrained system will be discussed in the following sections.

A system of quantum nonlinear oscillators is given by the following Hamiltonian:
\begin{eqnarray}\label{e:Ham}
\hat H&=&\sum_{i=1}^n\frac{1}{2m_i}\hat P_i^2+V(\hat Q_1,\hat Q_2,\dots,
\hat Q_n)\nonumber\\
&=& \sum_{i=1}^n\frac{1}{2m_i}\hat P_i^2+\frac{m_i\omega_i^2}{2}
{\hat Q_i}^2+\dots,
\end{eqnarray}
where $V$ is some function of $(\hat Q_1, \hat Q_2,\dots,\hat Q_n)$ having
the properties $\partial^2 V/\partial{Q_i^2}|_{Q_i=0}=m_i\omega_i^2$
$(i=1,2,\dots,n)$.

In general case when the Hamiltonian is not only quadratic in
$\hat Q_1,\hat Q_2,\dots,\hat Q_n$, the dispersions $\Delta Q_i$,
$\Delta P_i$ ($i=1,2,\dots,n$) will assume different arbitrary high
values in the states along an orbit generated by $\hat H$.
However, the constrained system defined by the Hamiltonian (\ref{e:Ham})
and the following set of $2n$ constraints
\begin{subequations}\label{e:Disp}
\begin{align}
f^i_q(X)&=(\Delta\hat Q_i)^2-\frac{1}{2m_i\omega_i}=0,\\
f^i_p(X)&=(\Delta\hat P_i)^2-\frac{m_i\omega_i}{2}=0,
\end{align}
\end{subequations}
should preserve the dispersions of all fundamental quantum
observables. The values of the dispersions in (\ref{e:Disp}) are
the minimal values that can be achieved simultaneously by the
coordinates and momenta, and are obtained if and only if the state
of the $i$-th oscillator is a coherent state. However, the
constraints (\ref{e:Disp}) are irregular and we shall see that the
conservation of minimal dispersions is achieved by a more suitable
set of constraints.

Let us consider in detail a single nonlinear oscillator. This
example is in fact sufficient to indicate the typical features of
the general case. In this case there are only two constraints of
the form (\ref{e:Disp})
\begin{subequations}\label{e:Disp1}
\begin{align}
f_q(X)&=(\Delta\hat Q)^2-\frac{1}{2m\omega}=0,\\
f_p(X)&=(\Delta\hat P)^2-\frac{m\omega}{2}=0.
\end{align}
\end{subequations}
The constrained manifold $\Gamma$ defined by (\ref{e:Disp1})
coincides with the set of coherent states, which is a
finite-dimensional submanifold of all quantum states ${\cal M}$.
This shows that there exists an infinite set of constraints on
${\cal M}$ with the same constrained manifold as the one given by
the two constraints (\ref{e:Disp1}). Furthermore, this indicates
that it might not be possible to treat the two constraints
(\ref{e:Disp1}) within the standard Dirac scheme for regular
constraints. Nevertheless, in order to illustrate the problems
that occur, we shall proceed with the analysis of the constrained
Hamiltonian equations (\ref{e:Xdot}) with the two constraints
(\ref{e:Disp1}).

The general dynamical equations for the fundamental variables
$q=\langle\hat Q\rangle$, $p=\langle\hat P\rangle$ of the constrained quantum Hamiltonian system with the constraints (\ref{e:Disp1}) assume the form
\begin{eqnarray}\label{e:qp}
\dot q&=&\{q,H+\lambda_qf_q+\lambda_pf_p\}_{\cal M},\nonumber\\
\dot p&=&\{p,H+\lambda_qf_q+\lambda_pf_p\}_{\cal M},
\end{eqnarray}
and should be solved together with the constraints equations (\ref{e:Disp1}).
Notice that in (\ref{e:qp}) the Poisson brackets are that of the
full quantum phase space ${\cal M}$, and $H$ is a function on $\cal M$
and not on the constrained manifold.

The general procedure requires first to compute the Poisson brackets between the constraints and between the constraints and the Hamiltonian $H$, and then check
the values they assume on the constrained manifold $\Gamma$. Computations are facilitated using the relations
\begin{subequations}
\begin{eqnarray}
{\delta\over \delta\psi(q)}\langle\psi|\hat A|\psi\rangle&=
\langle\psi|\hat A|q\rangle,\\
{\delta\over \delta\psi^*(q)}\langle\psi|\hat A|\psi\rangle&=
\langle q|\hat A|\psi\rangle,
\end{eqnarray}
\end{subequations}
and equality
\begin{align}
\begin{split}
&{\delta A_1\over \delta\phi(q)}{\delta A_2\over \delta\pi(q)}-
{\delta A_2\over \delta\phi(q)}{\delta A_1\over \delta\pi(q)}=\\
&-i\left[{\delta A_1\over \delta\psi(q)}{\delta A_2\over \delta\psi^*(q)}-
{\delta A_2\over \delta\psi(q)}{\delta A_1\over \delta\psi^*(q)}\right],
\end{split}
\end{align}
where $\psi^*(q)=(\phi(q)-i\>\!\pi(q))/\sqrt{2}$.

The Poisson brackets between the constraints are
\begin{align}\label{e:fqfp}
&\{f_q,f_p\} = -i\int dq \left[
{\delta f_q\over \delta \psi(q)}{\delta f_p\over \delta\psi^*(q)}
-{\delta f_p\over \delta\psi(q)}{\delta f_q\over \delta\psi^*(q)}
\right]\nonumber\\
&= -i\int dq \left[
\big(\langle\psi|\hat Q^2|q\rangle-2\langle\hat Q\rangle
\langle\psi|\hat Q|q\rangle\big)\cdot\big(\langle q|\hat P^2|\psi\rangle
\right.\nonumber\\
&\left.-2\langle\hat P\rangle\langle q|\hat P|\psi\rangle\big)-\leftrightarrow
\right]=-i\left(\langle[\hat Q^2,\hat P^2]\rangle-2\langle\hat Q\rangle
\langle[\hat Q,\hat P^2]\rangle\right.\nonumber\\
&-\left. 2\langle\hat P\rangle\langle[\hat Q^2,\hat P]\rangle+4\langle\hat Q\rangle\langle\hat P\rangle\langle[\hat Q,\hat P]\rangle\right)\nonumber\\
&=2\Big(\langle\hat Q\hat P+\hat P\hat Q\rangle-2\langle\hat Q\rangle
\langle\hat P\rangle\Big)\equiv 4\,\Delta(\hat Q,\hat P).
\end{align}
The symbol $\leftrightarrow$ means the term of the same form as the
previous one but having $\hat Q$ replaced by $\hat P$ and vice versa.

Similar calculations give the brackets between the constraints and the Hamiltonian $H$
\begin{equation}\label{e:fqH}
\{f_q,H\}=\frac{2}{m}\,\Delta(\hat Q,\hat P),
\end{equation}
\begin{equation}\label{e:fpH}
\{f_p,H\}=-2\,\Delta(V'(\hat Q),\hat P),
\end{equation}
where $V'(\hat Q)$ denotes the derivative of the function $V(\hat Q)$.

All three expressions (\ref{e:fqfp}), (\ref{e:fqH}) and (\ref{e:fpH}) are zero
on the constrained manifold $\Gamma$ of the coherent states. Thus the
constraints (\ref{e:Disp1}) appear to be of the first class and there are
no secondary constraints. According to the general theory for the regular first
class constraints the Lagrange multipliers in the total Hamiltonian $H_{tot}$
should be left unspecified. However the constraints are not regular because,
for example, the derivative with respect to the coordinate $\psi(q)=\psi(0)$ is
\begin{equation}
{\delta f_q\over \delta\psi(q)}=\left[q^2\psi^*(q)
-2 q\langle\hat Q\rangle\psi^*(q)\right]\Big|_{q=0}=0,
\end{equation}
indicating that the multipliers have to be specified from some other condition.
In order to correctly fix the multipliers one might use the following reasoning. Consider the Poisson bracket $\{f_q,f_p\}$. If the constraints where regular,
this bracket would be a first class constraint and would be preserved by the
constrained evolution. In fact, the bracket $\{f_q,f_p\}$ computed on $\Gamma$,
as seen from (\ref{e:fqfp}), represents the correlation between $\hat Q$ and
$\hat P$ in a coherent state. This must be preserved by the evolution on $\Gamma$
generated by the total Hamiltonian $H_{tot}$. The dynamical equation with
$H_{tot}$ for the correlation $\Delta(\hat Q,\hat P)$ reads
\begin{align}\label{e:fqp_dot}
\frac{d}{dt}\Delta(\hat Q,\hat P)&=\{\Delta(\hat Q,\hat P),H_{tot}\}\nonumber\\
&=2\Big({1\over 2m}(\Delta\hat P)^2-{\langle V''(\hat Q)\rangle\over 2}
(\Delta\hat Q)^2\nonumber\\
&\qquad+\lambda_p(\Delta\hat P)^2-\lambda_q(\Delta\hat Q)^2\Big),
\end{align}
on $\Gamma$ and vanishing only if the multipliers are
\begin{equation}
\lambda_p=-{1\over 2m},\qquad \lambda_q=-{\langle V''(\hat Q)\rangle\over 2}.
\end{equation}
Thus, the total Hamiltonian that would preserve the irregular constraints (\ref{e:Disp1}) with the additional compatibility condition is
\begin{equation}\label{e:Htot1}
H_{tot}={\langle \hat P\rangle^2\over 2m}+\langle V(\hat Q)\rangle
-{\langle V''(\hat Q)\rangle\over 2}((\Delta\hat Q)^2-{1\over 2m\omega}).
\end{equation}
However, this is still not satisfactory. To see this, one might
observe that $\Delta(f(\hat Q),\hat P)=0$ should hold on $\Gamma$
for arbitrary $f(\hat Q)$. The evolution generated by the total
Hamiltonian (\ref{e:Xdot}) should yield on $\Gamma$
\begin{equation}\label{e:f(Q)P}
\frac{d}{dt}\Delta(f(\hat Q),\hat P)=\{\Delta(f(\hat Q),\hat P),H_{tot}\}=0.
\end{equation}
It turns out that the multiplier $\lambda_q$ must depend on $f(\hat Q)$
i.e.\ on {\it arbitrary} function and cannot be fixed by any means. The
origin of such discrepancy is seen from
\begin{align}\label{e:<V>}
\langle V(\hat Q)\rangle=\sum_{k=0}^{\nu}\frac{V^{(k)}(\langle \hat Q\rangle)}
{k!}\langle(\hat Q-\langle\hat Q\rangle)^k\rangle,
\end{align}
where possibly all moments $\langle(\hat Q-\langle\hat Q\rangle)^k\rangle$
are present (if $\nu=\infty$)  and {\it influence the dynamics} (\ref{e:f(Q)P}),
while our constraints (\ref{e:Disp1}) contain only the moment of order two.
Resolution of this problem requires number of constraints equal to the order
$\nu$ of highest moment present in (\ref{e:<V>})
\begin{subequations}\label{e:Disp2}
\begin{align}
&f_{q,2k-1}(X)=\langle(\hat Q-\langle\hat Q\rangle)^{2k-1}\rangle=0,\\
&f_{q,2k}(X)=\langle(\hat Q-\langle\hat Q\rangle)^{2k}\rangle-
\frac{(2k-1)!!}{(2m\omega)^k}=0,
\end{align}
\end{subequations}
$k=2,3,\dots,\lfloor(\nu+1)/2\rfloor.$
Although the constraints (\ref{e:Disp2}) implicitly follow from (\ref{e:Disp1})
and hold automatically on $\Gamma$, they must be present explicitly in total
Hamiltonian. In that case choice of the multipliers
\begin{equation}
\lambda_{q,k}=-\frac{V^{(k)}(\langle \hat Q\rangle)}{k!}\Big|_\Gamma,
\end{equation}
cancel term-wise the appropriate contributions of moments
$\langle(\hat Q-\langle\hat Q\rangle)^k\rangle$ to the evolution
(\ref{e:f(Q)P}).

We see that starting with the primary constraints (\ref{e:Disp1}) one
would have to add a possibly infinite number of secondary constraints in
order to satisfy all possible compatibility conditions (\ref{e:f(Q)P}). This
is not satisfactory. Fortunately, there is an alternative procedure
which starts with the different set of two primary constraints and offers
the resolution.

\subsection{More convenient primary constraints}

To formulate the primary constraints in the alternative procedure,
we associate with each point from ${\cal M}$ denoted $X_{\psi}$ a
point $\alpha(\psi)$ on the coherent state manifold $\Gamma$ such
that
\begin{equation}\label{e:alpha}
\alpha(\psi)=(\langle\hat Q\rangle_\psi,\langle\hat P\rangle_\psi).
\end{equation}
  By definition, the operators $\hat Q$ and $\hat P$
have the expectations in the coherent state $\alpha(\psi)$ the
same as in the state $\psi$. This association of a single coherent
state with the whole set of states in fact establishes an
equivalence relation on ${\cal M}$, that will play a crucial role
in the following section.

With the notation (\ref{e:alpha}) we formulate the following two constraints
\begin{subequations}\label{e:CONSTR}
\begin{align}
&\Phi_q=\langle V(\hat Q)\rangle_\psi-\langle V(\hat Q)\rangle_{\alpha(\psi)}=0,\\
&\Phi_p=\langle\hat P^2\rangle_\psi-\langle\hat P^2\rangle_{\alpha(\psi)}=0,
\end{align}
\end{subequations}
to be imposed on the oscillator with arbitrary fixed potential
$V(\hat Q)$.

The total Hamiltonian assumes the standard form
\begin{equation}
H_{tot}=\langle\hat H\rangle_{\psi}+\lambda_q\Phi_q+\lambda_p\Phi_p,
\end{equation}
and the compatibility condition
\begin{equation}
\{\Delta (f(\hat Q),\hat P),H_{tot}\}=0,
\end{equation}
yields the values of Lagrange multipliers
\begin{equation}
\lambda_q=-1,\qquad \lambda_p=-{1\over 2m},
\end{equation}
independently of the function $f(\hat Q)$, leading
to\begin{equation} H_{tot}={1\over 2m}\langle\hat
P^2\rangle_{\alpha(\psi)}+ \langle V(\hat
Q)\rangle_{\alpha(\psi)}\equiv\langle\hat H\rangle_{\alpha(\psi)}.
\end{equation}
Noting that $\langle\hat P^2\rangle_{\alpha(\psi)}=
\langle\hat P\rangle_{\alpha(\psi)}^2+m\omega/2$ and dropping irrelevant
constant we finally obtain the total constrained Hamiltonian
\begin{equation}\label{e:HTOT}
H_{tot}={1\over 2m}\langle\hat P\rangle_{\alpha(\psi)}^2+
\langle V(\hat Q)\rangle_{\alpha(\psi)}
\end{equation}
that preserves the evolution on the manifold of the coherent states $\Gamma$.

The important fact is that the total Hamiltonian (\ref{e:HTOT}) depends only
on the variables $q\equiv\langle\hat Q\rangle_\psi$ and
$p\equiv\langle\hat P\rangle_\psi$ that parametrize the coherent state manifold.
Furthermore, it is seen that the total Hamiltonian (\ref{e:HTOT}) is up to
additive constant equal to the initial Hamiltonian $H\equiv \langle\hat H\rangle
_{\psi}$ on the constrained manifold $\Gamma$. However, $H_{tot}$ preserves
constant and minimal quantum fluctuations of fundamental observables, while
the evolution with $H$ can in general make them quite large.

\subsection{Quantum constrained system and the classical oscillator}

We shall now compare the total Hamiltonian (\ref{e:HTOT}) on the
constrained manifold $\Gamma$ of the coherent states with
\begin{equation}
h_{cl}={1\over 2m}p^2+V(q).
\end{equation}
representing the Hamilton' s function of a classical nonlinear
oscillator with the potential $V(q)$.

The quantum expectation of the potential $V(\hat Q)$ in a coherent state
$\alpha$ is
\begin{equation}\label{e:<V>_cs}
\langle V(\hat Q)\rangle_{\alpha}=\int_{-\infty}^\infty V(x) {{\exp\left(-{(x-\langle\hat Q\rangle_\alpha)^2\over
2(\Delta\hat Q)^2_\alpha}\right)\over (\Delta\hat Q)_\alpha\sqrt{2\pi}}} dx.
\end{equation}
Using the general formula
\begin{equation}
\int_{-\infty}^{\infty} f(t){\exp\left(-{(x-t)^2\over 4q^2}\right)
\over 2q\sqrt{\pi}} dt = \sum_{k=0}^{\infty} {q^{2k}\over k!} f^{(2k)}(x)
\end{equation}
we see that
\begin{equation}
\langle V(\hat Q)\rangle_\alpha = V(q)+\sum_{k=1}^{\infty} {(\Delta Q)_\alpha
^{2k}\over 2^k k!}V^{(2k)}(q),
\end{equation}
where $q=\langle\hat Q\rangle_\alpha$ and $(\Delta
Q)_\alpha=1/\sqrt{2m\omega}$. Thus, the total Hamiltonian in a
point $\alpha$ on the constrained manifold is
\begin{eqnarray}\label{e:H_sum}
H_{tot}&=&{p^2\over 2m}+V(q)+\sum_{k=1}^{\infty} {1\over 2^k k!}{V^{(2k)}(q)\over (2m\omega)^k}\nonumber\\
&\equiv& h_{cl}+\sum_{k=1}^{\infty} {1\over 2^k k!}{V^{(2k)}(q)\over (2m\omega)^k}.
\end{eqnarray}
In the limit of large mass $m$ the terms in the sum in (\ref{e:H_sum})
approach zero yielding
\begin{equation}
H_{tot}\rightarrow h_{cl},\quad m\rightarrow\infty.
\end{equation}
Alternatively, the dispersion $(\Delta
Q)_\alpha=1/\sqrt{2m\omega}\rightarrow 0$ and the exponent in the
integral in (\ref{e:<V>_cs}) approaches the delta function
$\delta(x-\langle\hat Q\rangle_\alpha)\equiv\delta(x-q)$ producing
$\langle V(\hat Q)\rangle_\alpha\rightarrow V(q)$.

To summarize, we have formulated a consistent set of dynamical
equations for an arbitrary quantum nonlinear oscillator that
maintain the evolution on the coherent state manifold. Because
such evolution preserves minimal fluctuations $\Delta\hat Q$ and
$\Delta\hat P$, the total Hamiltonian $H_{tot}$ on $\Gamma$
differs from the Hamilton' s function of a classical nonlinear
oscillator with the same potential $V(q)$ by the terms that are
small for an oscillator of a macroscopic mass. At the risk of
repeating ourselves, let us stress once again that during the
evolution with the quantum Hamiltonian of the oscillator
$\langle\hat H \rangle_\psi$ with no constraints, the quantum
fluctuations $\Delta\hat Q$ and $\Delta\hat P$ can become large
and thus make Hamiltonian functions $\langle\hat H\rangle_\psi$
and $h_{cl}$ quite different even in the macroscopic limit.

For the system with more than one oscillators, that might be nonlinear
and interacting, the condition that $\Delta\hat Q_i$ and $\Delta\hat P_i$
are simultaneously minimal implies that each of the oscillators is always
in some pure $H_4$ coherent state $|\alpha_i(t)\rangle$. Thus, the total state
$|\psi(t)\rangle$ is always given by the tensor product of the single
oscillator's pure coherent states $|\psi(t)\rangle=\otimes_i
|\alpha_i(t)\rangle$, implying for example
\begin{eqnarray}
\langle\psi(t)|\hat Q_1\otimes\hat Q_2|\psi(t)\rangle &=&
\langle\hat Q_1\rangle_{\alpha_1(t)}\times
\langle\hat Q_2\rangle_{\alpha_2(t)}\nonumber\\
&=& q_1(t)\times q_2(t).
\end{eqnarray}

Suppression of quantum fluctuations for each oscillator's degree of
freedom implies that the degrees of freedom of different oscillators do
not get entangled during the evolution. This is enough to generalize
the results of the single oscillator analysis to the general case of
arbitrary number of interacting oscillators with constraints.

We have formulated the constrained evolution of a quantum system
of oscillators with the corresponding constraints. In general, the
Hilbert space of a quantum system represents the space of an
irreducible representation of the corresponding dynamical algebra
$\mathsf g$, that need not be the Heisenberg algebra as it is in the
case of oscillators. Nevertheless, one could study the evolution of
such a system with the constraints analogous to (\ref{e:Disp}). The
constraint manifold of such a system with a Lie dynamical algebra
$\mathsf g$ should coincide with the manifold of the corresponding
$\mathsf g-$generalized coherent states \cite{Perelomov,PhysA,Delburgo}.

\section{Equivalence relation among the quantum states}

The fundamental quantum observables $\hat Q_i$, $\hat P_i$
$(i=1,2,\dots,n)$ define $2n$ functions $\langle X|\hat
Q_i|X\rangle$, $\langle X|\hat P_i|X\rangle$ on $\cal M$. Values
that these functions take  on the coherent states, parameterize
the $2n-$dimensional manifold of the coherent states $\Gamma$.
Thus, the set of fundamental quantum observables and the
constrained manifold are seen to be in a one-to-one relation.

We use the coherent states or the elementary quantum observables
$\hat Q_i,\hat P_i$ to define an equivalence relation on ${\cal M}$.
Two general quantum states $X_1\in{\cal M}$ and $X_2\in{\cal M}$ are
defined to be equivalent, or {\it physically indistinguishable}, if
each fundamental quantum observable takes the same value in $X_1$
as in $X_2$. Thus, $X_1\sim X_2$ iff $q_i(X_1)=q_i(X_2)$,
$p_i(X_1)=p_i(X_2)$ $(i=1,2,\dots,n)$. An equivalent definition is
that the states $X_{1,2}$ are equivalent iff there is a coherent
state $(q,p)$ such that $q_i(X_{1,2})=q_i(q,p)=q_i$,
$p_i(X_{1,2})=p_i(q,p)=p_i$ $(i=1,2\dots,n)$. Each equivalence
class contains one and only one coherent state, i.e.\ a state from
the constraint manifold $\Gamma$.

The quantum phase space ${\cal M}$ appears as a bundle over the
constraint manifold $\Gamma = {\cal M}/\!\!\sim$. $\Gamma$ is even
dimensional, and is parameterized by the values of only $2n$
independent variables $(q_i,p_i$), $i=1,2,\dots,n$. $\Gamma$
inherits a symplectic structure $\omega$ which is the pull-back of
the symplectic structure $\Omega$ on ${\cal M}$. In fact $\Gamma$
is finite-dimensional symplectic manifold and $(q_i,p_i)$,
$i=1,2,\dots,n$ are canonical coordinates. Thus, the constraint
manifold $\Gamma$ is the phase space of a classical system of $n$
oscillators. This is the way in which the phase space of a
classical mechanical system appears from the structure of the
quantum mechanics.

We have seen that: a) the constrained manifold $\Gamma$ is related
to a certain equivalence relation on full quantum phase space ${\cal M}$
and b) $\Gamma$ has the phase-space structure of a finite Hamiltonian
dynamical system. We can distinguish two dynamical systems on $\Gamma$
defined by Hamilton's functions $H_{tot}$ restricted on $\Gamma$ and $h_{cl}$.
Since $\Delta\hat Q=(2m\omega)^{-1/2}=const$ during the evolution
defined by $H_{tot}$ such evolution differs from the dynamics generated
by $h_{cl}$ by the terms which are small in the macroscopic limit.

\section{Discussion}

The presented picture where the constraints are seen as the
equivalence relation imposed on the quantum states suggests a
physical interpretation of the constrained Hamiltonian system
$(\Gamma,\omega,H_{tot}|_{\Gamma})$. The equivalence classes of
quantum states determine the corresponding quantum observables
that can be considered as physically distinguishable. Thus, in the
Hamiltonian system with constraints only functions defined on
$\Gamma$ are considered as physically distinguishable. In other
words, if two functions on ${\cal M}$ correspond to two different
operators but generate the same function on $\Gamma$, the two
operators should be considered as physically indistinguishable. We
see that imposing the constraint on the quantum system in fact
provides the mathematical representation of a coarse-grained
description of the quantum system.

The coarse-grained description gives a system with the
kinematic properties of a classical Hamiltonian mechanical
system. Furthermore, dynamics of the constrained system is such
that the quantum fluctuations of fundamental observables
are constant and simultaneously minimal during the evolution.
In fact, one can identify a class of classical Hamiltonian
dynamical systems that is generated by the constrained quantum
system and that preserves the quantum fluctuations. The systems
in this class differ from each other by terms that are arbitrary
small for sufficiently large value of the masses. On the contrary,
the corresponding terms in the quantum Hamiltonian system with no
constraints, i.e.\ in the full-detail picture without the
coarse-graining, necessarily become large during the evolution.
They are responsible for the creation of typically quantum
superpositions.

It is well known that a generic Hamiltonian dynamical system is
not structurally stable, i.e.\ small perturbations of the
Hamilton's function typically induce non-equivalent phase portraits
\cite{Arnold,Arnold1,PhysD}. Thus, one can expect qualitative
differences between the quantum systems for large values of the
classicality parameter and the classical system. However, this is
the problem of any Hamiltonian theory as a framework for robust
modelling of dynamical phenomena, and is not strictly related to
the QC-relation.

We see that the classical system appears because of: a) the
coarse-grained description of the quantum system and then b) in
the macroscopic limit corresponding to the large masses. It is
important to note that the two factors, i.e.\ the coarse-graining
and the macro-limit, are independent and both are necessary
(please see also \cite{Brukner1,Brukner2}). The two factors, one
leading to the suppression of, i.e.\ impossibility to observe,
dynamically created quantum coherences and the other involving the
macro-limit also appear in other explanations of the appearance of
the classical world from the quantum, like for example in the
theory of environmentally induced decoherence
\cite{decoh1,decoh2}.

Finally, let us illustrate the independent roles played by the
macro-limit and the coarse-grained observation using one more example.
Consider a large collection of $1/2$ spins $\hat\sigma^i$, $i=1,2,\dots,N$.
One can define collective quantum observables $\hat m_x=\sum_i\hat\sigma^i_x/N$,
$\hat m_y=\sum_i\hat\sigma^i_y/N$, $\hat m_z=\sum_i\hat\sigma^i_z/N$.
The macro-limit corresponds in this case to the limit of large $N$.
However, the macroscopic magnetizations $m_{x,y,z}=\langle\hat m_{x,y,z}\rangle$
in general do not behave as classical variables. Even if the initial state
is such that $\Delta m_{x}/m_x$, $\Delta m_{y}/m_y$ and $\Delta m_{z}/m_z$
are all small, the evolution might be such that quite quickly these ratios
become large i.e.\ close to unity \cite{jaPRA}. This occurs if the Hamiltonian
includes long range interactions, for example if $H_{int}=\sum_{i,j}\hat\sigma^i\hat\sigma^j$. Thus, the macro-limit alone
does not imply the classical behaviour even for the selected set of
global observables. This has been nicely illustrated in \cite{Brukner1,Brukner2} (please see also \cite{jaPRA}). A coarse-graining analogous to the one
discussed in this paper is also needed. One declares that the only
states that are physically measurable are necessarily such that
$\Delta m_{x}$, $\Delta m_{y}$, $\Delta m_{z}$ are simultaneously minimal.
The states satisfying this condition are the $SU(2)$ coherent states of
the $N$-term direct product representation. Such coarse-graining is
equivalent to the evolution constrained on the submanifold of these
coherent states so that all three dispersions are small during
such evolution. Notice that the coarse-graining also implies that
the eigenstates of the quantum collective variables $\hat m_{x,y,z}$
are not among the physically measurable states. Equally, the states
corresponding to a superposition of states with very different
values of the macroscopic variables $\langle\hat m_{x,y,z}\rangle$
are not physical.

\section{Summary}

We have used the formulation of quantum dynamics in the form of a
Hamiltonian dynamical system to study the relation between quantum
and classical systems of nonlinear interacting oscillators. The
classical system has finite dimensional phase-space and the
quantum system viewed as the Hamiltonian system is infinite
dimensional in an essential way. Kinematical and dynamical
properties of the classical system are obtained from the quantum
one via the two step procedure consisting of: a) coarse-graining
and b) macroscopic limit. The coarse-graining is mathematically
treated as an equivalence relation on the set of quantum states,
and as a result  emerges the classical phase-space. The
equivalence relation imposes a constraint on the Hamiltonian
dynamics of the quantum system. The effect of the constraints is
to preserve constant and minimal quantum fluctuations of the
canonical observables. The formulation of the most appropriate
finite set of constraints that fulfill the goal is not
straightforward, and involves the nonlinear potential. Resulting
constrained Hamiltonian system on the constrained manifold
represents the coarse-grained description of the quantum system of
oscillators. The system differs from the classical system with the
same potential only in the terms that are arbitrary small for
oscillators with sufficiently large mass, i.e. in the macroscopic
limit.

The procedure can be generalized to obtain other classical systems
from the corresponding coarse-grained quantum systems in the
corresponding macroscopic limit.

\begin{acknowledgments}
This work is partly supported by the Serbian Ministry of Science contracts No.\
171017, 171028, 171038 and 45016.
\end{acknowledgments}

\end{document}